\documentclass[11pt,onecolumn,draftcls,english]{IEEEtran}

\usepackage{color,slashbox}
\usepackage{graphicx}
\usepackage{amsmath, dsfont}
\usepackage{amsfonts}
\usepackage{amssymb}
\usepackage{xspace}
\usepackage{array}
\usepackage{epsfig}
\usepackage{float}
\usepackage{soul, pgfplots}		         % utilisÈ dans les macros de Charles

\DeclareMathOperator{\E}{E}

\DeclareMathOperator{\diag}{diag}
\DeclareMathOperator*{\argmin}{arg\,min}
\DeclareMathOperator*{\argmax}{arg\,max}

\newcommand{\eg}{\emph{e.g., }}
\newcommand{\ie}{\emph{i.e., }}

\newcommand{\Rr}{\mathds{R}}

\newcommand{\D}{\mathbf{D}}

\newcommand{\I}{\mathbf{I}}

\newcommand{\thev}{\text{\boldmath$\theta$}}

\newcommand{\etav}{\text{\boldmath$\eta$}}

\newcommand{\yv}{\mathbf{y}}

\newcommand{\dv}{\mathbf{d}}

\newcommand{\mv}{\mathbf{m}}

\newcommand{\rv}{\mathbf{r}}
\newcommand{\zv}{\mathbf{z}}

\newcommand{\tv}{\text{\boldmath$\theta$}}

\newcommand{\Ns}{\mathcal{N}}

\def\Pn{{\mathbf{P}}}
\def\zv{{\mathbf{z}}}
\def\noise{{\boldsymbol{\omega}}}
\def\obs{{\mathbf{y}}}

\title{DOA estimation in structured phase-noisy environments: technical report}%:\\Bayesian formulation and sparse priors}
\author{\IEEEauthorblockN{Ang\'elique Dr\'emeau,\thanks{A. Dr\'emeau is with ENSTA Bretagne and Lab-STICC UMR 6285, France. C. Herzet is with INRIA Centre Rennes - Bretagne Atlantique, France, and is currently working in Lab-STICC UMR 6285, France.} and C\'edric Herzet}}
\begin{document}
%\ninept
%
\maketitle
\begin{abstract}
In this paper we focus on the problem of estimating the directions of arrival (DOA) of a set of incident plane waves. 
%This paper deals with the estimation of the directions of arrival (DOA) of a set of incident plane waves. 
Unlike many previous works,  which assume that the received observations are only affected by additive noise, we consider the setup where some phase noise also corrupts the data (as for example observed in atmospheric sound propagation or underwater acoustics). We propose a new methodology to solve this problem in a Bayesian framework by resorting to a variational mean-field approximation. Our simulation results illustrate the benefits of carefully accounting for the phase noise in the DOA estimation process. 
%the waves travel through highly fluctuating media
%Unlike the standard setup, with typically suppose that the received signal is only corrupted by additive noise, 
%Most practical cases of direction-of-arrival estimation often involve fluctuating media.
\end{abstract}

\section{Introduction}
\label{sec:intro}
 Estimating the directions of arrival (DOA) of propagating plane waves is at the heart of many applicative domains including sonar, radar and mobile telecommunications. Among the rich literature dealing with this problem, the most popular method is probably conventional beamforming \cite{Johnson1993} which can roughly be interpreted as a least-square estimator. 
%In practice, the approach presents good performance when considering long antenna. 
\textit{Per~se}, this approach suffers from a lower limit on the resolution achievable in the DOA estimation process (conditioned by the length of the sensor array).~To overcome this issue, so-called ``high-resolution'' techniques, taking advantage of more prior information on (the number and the nature of)  the sources, have been proposed in the literature (see \eg \cite{Schmidt1986,Mantzel2012, Xenaki2014}). In \cite{Schmidt1986}, the authors introduced the well-known MUSIC algorithm which benefits from the knowledge of the number of the sources and rely on the assumption that the noise and the signal of interest live in perfectly separable subspaces. More recently, a ``compressive'' beamforming approach was proposed in \cite{Mantzel2012, Xenaki2014}, where a sparse prior was exploited to address the DOA estimation problem.

The contributions mentioned above assume that the incident plane waves are only corrupted by some additive noise.~Unfortunately, when the waves travel through highly fluctuating media, as in the case of \eg atmospheric sound propagation \cite{Cheinet2012} or underwater acoustics \cite{Flatte2010}, this model does no longer describe accurately the physics underlying the propagation process.~In such cases,
% as emphasized \eg in \cite{Cheinet2012,Flatte2010} \remCH{a verifier $\leadsto$ AD: non, les ref \cite{Cheinet2012, Flatte2010} mettent juste en avant le pb et tentent des explications purement physiques, je retirerais donc le ``as emphasized in"}, 
a multiplicative phase noise typically corrupts the collected signal, making the corresponding DOA estimation problem quite challenging. We address this problem in the present paper. 

Our approach is inspired from the recent standard ``high-resolution'' DOA methods \cite{Mantzel2012, Xenaki2014} and some phase retrieval algorithms presented in \cite{Schniter2012, Dremeau2015}. More specifically, we model the received signal as a sparse combination of elementary signals  (taken from a redundant dictionary) and assume that the latter is corrupted by both additive and phase noise. Our methodology is grounded on a probabilistic Bayesian framework and relies on a variational mean-field approximation.
% \remCH{Plutot inclure la phrase qui suit dans la discussion a la fin de la section 2?} 
In particular, we show how to nicely incorporate fine noise-phase models in this framework, extending in this respect, the approaches proposed in \cite{Schniter2012, Dremeau2015}.

\section{Problem Statement}
\label{sec:bayes}

%\subsection{Observation model}
%\label{subsec:model}
%\remCH{A discuter...}

Our derivations are based on the following formulation of the DOA estimation problem: we consider an antenna composed of $N$ sensors and 
assume that the collected observation vector $\mathbf{y}\in\mathbb{C}^N$ can be expressed as %obeys the following model
\begin{align}
\mathbf{y} &= \Pn\D \zv + \noise,\label{eq:obs}
\end{align}
where $\noise\in\mathbb{C}^{N}$ and $\Pn=\diag(\lbrace e^{j\theta_n}\rbrace_{n=1}^{N})\in\mathbb{C}^{N\times N}$ play respectively the role of an additive and a multiplicative phase  noise. Matrix $\D=[\dv_1\ldots\dv_M]\in\mathbb{C}^{N\times M}$ is made up of the steering vectors $\dv_i \triangleq [e^{j\frac{2\pi}{\lambda} \Delta \sin(\phi_i)}\ldots e^{j\frac{2\pi}{\lambda} \Delta N \sin(\phi_i)}]^T$,  where $\phi_i$'s are some potential angles of arrival, $\Delta$ is the distance between two adjacent sensors, and $\lambda$ is the wavelength of the propagation waves. 

With this formulation, assuming that $\obs$ results from the combination of a few waves arriving from different angles $\phi_i$'s, the DOA estimation problem is basically equivalent to identifying the positions of the nonzero coefficients in $\zv$ (since each column of $\D$ corresponds to a particular angle of arrival).~In the standard DOA estimation problem, $\Pn$ is assumed to be known with $\Pn=\I_N$, where $\I_N$ is the $N\times N$ identity matrix. In this case, the model connecting the unknown vector $\zv$ to the measurements $\obs$ is linear; finding the position of the nonzero elements in $\zv$ can then be carried out with standard sparse-representation algorithms, see \eg~\cite{Foucart2013Mathematical}.

In this paper, we consider the more complex case where $\Pn=\diag(\lbrace e^{j\theta_n}\rbrace_{n=1}^{N})$ and the $\theta_n$'s are unknown. 
 More specifically, we assume that the phases $\theta_n$'s obey the following  Markov model:
\begin{align}
p(\thev) = \prod_{n = 2}^N p(\theta_n|\theta_{n-1})\;p(\theta_1),\label{eq:phaseprior}
\end{align}
with $p(\theta_n|\theta_{n-1}) = \Ns(a\,\theta_{n-1},\sigma_\theta^2)$, $\forall n\in\lbrace 2,\ldots,M\rbrace$, $a\in\mathbb{R_+}$, and $p(\theta_1) = \Ns(0,\sigma_1^2)$.
 From a practical point of view, this model allows us to describe spatial fluctuations of the propagation medium all along the antenna; the strength of the fluctuations is related to the value of parameter $\sigma_\theta^2$. 

As noted in the introduction, assuming that $\Pn$ is unknown renders the DOA estimation much more difficult since it introduces uncertainties on the observation model. 
%the joint estimation of $\zv$ and $\boldsymbol{\theta}$ is a highly nonlinear problem. 
Before proceeding to the presentation of the proposed methodology to address this issue, we draw some connections with other applicative fields. 
%
% First, our work relates to the phase retrieval problem (\eg \cite{candes2013phaselift}) where the phase information of the observations is completely missing: only intensities or amplitudes are acquired. Formally, both problems share - explicitely or not - the same observation model \eqref{eq:obs} but differ in the prior distribution they enforce on the phases $\tv$, the absence of phase information being modeled by a non-informative prior, such as a uniform law (see \cite{Schniter2012,Dremeau2015}). 
% %
% Second, we note that, in top of being relevant for the DOA estimation problem with fluctuating media, this model is also of interest in the domain of digital communications where it can be used to characterize the transmission of complex modulation symbols over a channel affected by carrier phase noise, see \eg  \cite{Colavolpe2005}. 

 \begin{itemize}
%In the literature, model (2) has been already considered in phase retrieval problems with a uniform prior on the phases ? (see e.g. [8, 9]). 
%In particular, our work relates to previous work \cite{Dremeau2015} 
\item Our work relates first to the phase retrieval problem (\eg \cite{candes2013phaselift}) where the phase information of the observations is completely missing: only intensities or amplitudes are acquired. Formally, both problems share - explicitely or not - the same observation model \eqref{eq:obs} but differ in the prior distribution they enforce on the phases $\tv$, the absence of phase information being modeled by a non-informative prior, such as a uniform law (see \cite{Schniter2012,Dremeau2015}). %\remAD{Pas d'accord avec le fait de dire que c'est generique: on se place quand meme dans un cas particulier ici, celui des chaines de markov, je m'arrete donc la sur ce point.}
\item Then we note that, in top of being relevant for the DOA estimation problem with fluctuating media, this model is also of interest in the domain of digital communications where it can be used to characterize the transmission of complex modulation symbols over a channel affected by carrier phase noise, see \eg  \cite{Colavolpe2005}. 
\end{itemize}
 
%\remCH{A mentionner: 
%\begin{itemize}
%\item Lien avec le probleme general du phase retrieval, qui considere implicitement que 
%\begin{align}
%p(\boldsymbol{\theta})= uniform.
%\end{align}
%%\item It is easy to see that if $\Pn=\I$, one exactly recovers DOA model as considered in \cite{Mantzel2012, Xenaki2014}. In this case, the model relating the unknown vector $\sv$ to the measurement $\obs$ is linear. Recovering a the target vector $\sv$ from $\obs$ by exploiting the sparsity of the former can then be done by standard methodologies, see \eg \cite{aaa}. 
%The algorithmic procedure derived in this paper is therefore also of interest from a methodological point of view to address general version of the phase retrieval problem. (bofement dit...)
%\item We note that, in top of being relevant for the DOA problem with fluctuating media, this model is also of interest in the domain of digital communications where it can used to characterize the transmission of complex modulation symbols over a channel affected by carrier phase noise, see \eg  \cite{Colavolpe2005}. 
%\end{itemize}
%}

\section{Bayesian Formulation of the Problem}

We address the problem of estimating $\zv$ from $\obs$ when the realizations of $\noise$ and $\boldsymbol{\theta}$ are unknown. To that end, we first place this problem into a Bayesian framework by defining suitable additional prior distributions on the unknown quantities.  

%First, the sparsity of  $\zv$ is accounted for via the following Bernoulli-Gaussian model: 
%
%We consider an antenna composed of $N$ sensors. $\forall n\in \lbrace 1,\ldots N\rbrace$, the measured acoustical field can then be written as, 
%%VERIFIER LA MULTIPLICATION DU BRUIT DE PHASE (SUR LE BRUIT)
%\begin{align}
%y_n 
%%&= e^{j\theta_n}\; z_n + \omega_n,\\
%&=e^{j\theta_n} \left(\sum_{i=1}^M z_i\, d_{n i} +\omega_n\right),\label{eq:obs}
%\end{align}
%where $\theta_n\in[0,2\pi)$ stands for the phase noise due to the fluctuations of the propagating medium and $\omega_n$ is a zero-mean circular Gaussian noise with variance $\sigma^2$. 
%
%In the context of DOA estimation, we assume that the dictionary $\D=[\dv_1\ldots\dv_M]$ is made up of so-called steering vectors such as $\dv_i = [e^{j\frac{2\pi}{\lambda} \Delta \sin(\phi_i)}\ldots e^{j\frac{2\pi}{\lambda} \Delta(L-1) \sin(\phi_i)}]^T$, where $\phi_i$ is the $i$-th DOA of a given set, $\Delta$ is the distance between two adjacent elements, and $\lambda$ is the assumed wavelength of the propagation wave.
%
%%Finally, the variables $x_i$, which weight the contributions of each direction of arrival in the measured acoustical field, are assumed to obey a zero-mean circular Gaussian law of variance $\sigma_x^2$.
%
%\subsection{Sparse prior on the number of sources}
%\label{subsec:sparseprior}
%

To account for the sparsity of $\zv$, we suppose that, $\forall i\in \lbrace1,\ldots, M\rbrace$, $z_i = x_i\cdot s_i$ with
\begin{align}
%p(y_n|\sv,\xv,\theta_n)=\mathcal{C}\Ns\left(e^{j\theta_n} \sum_{i=1}^M s_i\; x_i\; d_{n i},\sigma^2\right),\label{eq:beginmodel}\\
p(x_i)=\mathcal{C}\Ns(0,\sigma_x^2),\quad\text{and}\quad
p(s_i)=\text{Ber}(p_i).\label{eq:priors}
%\forall n\in \lbrace 1,\ldots N\rbrace\quad\quad\quad 
%p(\theta_n)&=\frac{1}{2\pi}.
\end{align}
This so-called Bernoulli-Gaussian model has been now largely used in the literature to model sparse priors (see \eg \cite{Dremeau2012}).
 Next, we assume that $p(\noise)$ is a zero-mean Gaussian distribution with variance $\sigma^2$. This hypothesis is typically justified by the central limit theorem under the assumption that the additive noise corrupting the data results from the aggregation of a large number of random parasitic contributions.
 Finally, the probabilistic model describing the behavior of $\boldsymbol{\theta}$ is given by \eqref{eq:phaseprior}. As mentioned in the previous section, this simple model accounts for local variations of the propagation medium along the sensor network. 

%Estimating the DOA from the dictionary $\D$ amounts to finding the nonzero coefficients in $\zv$.
%Based on this probabilistic model, our goal is to find a tractable approximation of the following estimator
%Exploiting model \eqref{eq:obs}-\eqref{eq:priors}, 
Based on this probabilistic model, we propose to look for the solution of the following Minimum Mean Square Error (MMSE) problem
%Maximum A Posteriori problem
\begin{align}
%z_i^\star&=\argmax_{z_i} p(z_i|\yv), \mbox{ $\forall i\in \lbrace1,\ldots, M\rbrace$},\label{eq:pb}
\hat{\zv} = &\argmin_{\tilde{\zv}} \E_{\zv|\yv}\left[\,\|\zv-\tilde{\zv}\|^2_2\right],\label{eq:pb}
\end{align}
relying on the marginal posterior distribution 
%$p(z_i|\yv) = \int_{\lbrace z_j\rbrace_{j\neq i}}\int_\tv\; p(\xv,\sv,\tv|\yv)\; d\xv\; d\tv$.
$p(\zv|\yv) = \int_\tv\; p(\zv,\tv|\yv)\; d\tv$. 
%
%With words, problem \eqref{eq:pb} aims at recovering the positions of the nonzero coefficients in $\zv$. 
% We note that, as long as the considered probabilistic model is the true generating model of the data, the estimator \eqref{eq:pb} is optimal in the sense that it minimizes the probability of making a wrong decision on each $s_i$. 
The computation of this marginalization being an intractable problem, we propose hereafter a practical procedure based on a mean-field approximation of the joint distribution 
 %$p(\mathbf{x}, \mathbf{s},\boldsymbol{\theta}\vert \obs)$ to approach the solution of \eqref{eq:pb}. 
 $p(\zv,\boldsymbol{\theta}\vert \obs)$ to approach the solution of \eqref{eq:pb}.

\section{The Proposed procedure}
\label{sec:procedure}

Mean-field approximations aim at approximating a posterior joint distribution by another one constrained to have a ``simple'' factorization while minimizing some distance with the targeted distribution. 
In the following, we will look for an approximation of $p(\zv,\boldsymbol{\theta}\vert \obs)$, say $\hat{q}(\zv,\tv)$, obeying the following factorization $\hat{q}(\zv,\tv) = \hat{q}(\tv) \prod_{i=1}^M \hat{q}(z_i)$. With this approximation of $p(\zv,\boldsymbol{\theta}\vert \obs)$, the evaluation of the marginal with respect to $\zv$ is simplified since $\int_\tv \hat{q}(\tv,\zv) = \prod_i \hat{q}(z_i)$ and the solution of the MMSE problem \eqref{eq:pb} can be approximated component-wise as \begin{align}\hat{z}_i \simeq \int_{z_i} z_i \; \hat{q}(z_i) \,dz_i.\label{eq:estiz}\end{align}

A well-known algorithm to find a mean-field approximation of a target distribution
% with a given factorization
 is the so-called ``Variational Bayes Expectation-Maximization'' (VBEM) algorithm (see \eg \cite{Beal2003}). Particularized to our problem, this procedure searches for a local minimum of the following optimization problem:
%Particularized to our problem, we propose to consider in this paper the following mean-field approximation $\hat{q}(\zv,\tv)$: %$\hat{q}(\xv,\sv,\tv)$:
\begin{align}
\hat{q}(\zv&,\tv)\! =\! \argmin_{q}\! \int_{\zv,\tv}\! q(\zv,\tv)\! \log\!\left(\!\frac{q(\zv,\tv)}{p(\zv,\tv|\yv)}\!\right)\!d\zv \,d\tv\quad
\text{ subject to } \quad q(\zv,\tv) = q(\tv) \prod_{i=1}^M q(z_i),\label{eq:MF}
%\hat{q}(\xv,&\sv,\tv)\! =\! \argmin_{q}\! \sum_\sv\! \int_{\xv,\tv}\! q(\xv,\sv,\tv)\! \log\!\left(\!\frac{q(\xv,\sv,\tv)}{p(\xv,\sv,\tv|\yv)}\!\right)\!d\xv d\tv\nonumber\\
%& \text{subject to } q(\xv,\sv,\tv) = q(\tv) \prod_{i=1}^M q(s_i,x_i).\label{eq:MF}
\end{align}
by sequentially minimizing the cost function with respect to $q(\tv)$ and $q(z_i)$, $\forall i\in\lbrace 1,\ldots,M\rbrace$. 

We show hereafter that the sequence of distributions generated by the VBEM algorithm, say $$\lbrace q^{(k)}(\tv), \lbrace q^{(k)}(z_i)\rbrace_i\rbrace_k,$$ admit simple parametric expressions.
%%\remCH{Il faut aussi dire que $q(\tv)$ et $q(z_i)$ sont des distribution de proba.}
%%\remCH{Il faut aussi dire rapidement pourquoi cela facilite le calcul de la marginale $q(s_i)$?}
%%Here, we propose to consider the so-called mean-field approximation such as
%%\begin{align}
%%\hat{q}(\xv,\sv,\tv|\yv) = \prod_{n=1}^N q(\theta_n)\prod_{i=1}^M q(x_i,s_i).
%%\end{align}
%A local minimum of this optimization problem can be found efficiently by means of the so-called Variational Bayes Expectation-Maximization (VBEM) algorithm (see \eg \cite{Beal2003}). This algorithm proceeds with a sequential minimization by updating each factor of the approximation one after the other.
%
%Particularized to model \eqref{eq:obs}-\eqref{eq:priors}, the procedure \addCH{will iterate over the update} of $q(\tv)$ and 
%%$q(x_i,s_i)$, $\forall i\in\lbrace 1,\ldots,M\rbrace$, 
%$q(z_i)$, $\forall i\in\lbrace 1,\ldots,M\rbrace$, 
%giving raise to a sequence of 
%%$\lbrace q^{(k)}(\tv), q^{(k)}(x_i,s_i)\rbrace_k$. 
%$\lbrace q^{(k)}(\tv), \lbrace q^{(k)}(z_i)\rbrace_i\rbrace_k$. 
%At each iteration, we show that the updated distributions have simple parametric expressions. 
% In the following subsections, we detail \addCH{the expression of $q^{(k)}(\tv)$ and $\lbrace q^{(k)}(z_i)\rbrace_i$} at a given iteration of the procedure. 
We detail these expressions (at a given iteration of the procedure) in the following subsections.
%Due to space limitation, we do not provide the technical derivations but refer the reader to our companion report \cite{DremeauTR} for more details.
 For the sake of clarity, we omit the iteration index $^{(k)}$ in the notations.

\subsection{Update of $q(\tv)$}
%Particularized to model \eqref{eq:obs}-\eqref{eq:phaseprior}, %the algorithm iterates between\footnote{For a sake of clarity, we omit the iteration index.}
%the procedure gives raise to a Gaussian posterior distribution for $\tv$:% $\theta_n$, $\forall n\in\lbrace 1,\ldots,N\rbrace$, 
%At a given iteration of the procedure, the estimate $q(\tv)$ writes as follows
The estimate of $q(\tv)$ is written as
\begin{align}
q(\tv)
&\propto \exp\left(\int_{\zv} \prod_i q(z_i) \log p(\yv,\zv,\thev)\; d\zv\right),\\
&\propto p(\tv)\;\exp\left(\int_{\zv} \prod_i q(z_i) \log p(\yv|\zv,\thev)\; d\zv\right),
\end{align}
with
\begin{align}
\log p(\yv|\zv,\tv)
&=\sum_{n=1}^N \log p(y_n|\zv,\theta_n),\\
&\propto \sum_{n=1}^N -\frac{1}{\sigma^2} \left( y_n^*y_n+\sum_{i=1}^M\sum_{k=1}^M z_i^*z_k\; d_{n i}^*d_{nk}  - 2 \;\Re\left\{y_n\left(\sum_{i=1}^M z_i^* d_{n i}^*\right) e^{-j\theta_n}\right\}\right),
%&\propto \sum_n -\frac{1}{\sigma^2} \left( y_n^*y_n+\sum_i\sum_k s_i s_k x_i^*x_k d_{n i}^*d_{n k} - e^{-j\theta_n}y_n \sum_i s_i x_i^*d_{n i}^* - e^{j\theta_n}y_n^* \sum_i s_i x_id_{n i} \right),
%&\propto \sum_n -\frac{1}{\sigma^2} \left( y_n^*y_n+\sum_i\sum_k s_i s_k x_i x_k d_{n i}^*d_{n k} - e^{-j\theta_n}y_n \sum_i s_i x_i d_{n i}^* - e^{j\theta_n}y_n^* \sum_i s_i x_id_{n i} \right),
\end{align}
where $\Re\{\cdot\}$ stands for the real part and $\cdot^*$ for the complex conjugate.% and $\eta_n \triangleq y_n\sum_{i=1}^M \langle z_i\rangle^*d_{ni}^*$.

Then, introducing $\eta_n \triangleq y_n\sum_{i=1}^M \langle z_i\rangle^*d_{ni}^*$, where $\langle z_i\rangle$ is defined using the current estimate of $q(z_i)$ as
% \suppCH{the current estimates $q(s_i)$ and} $m_{x_i}\!(s_i) = \int_{x_i} x_i\, q(x_i|s_i)$. \remCH{$q(s_i)$ et $q(x_i|s_i)$ non definis a ce stade: a deplacer plus haut?}
\begin{align}
 \langle z_i\rangle
&\triangleq \int_{z_i} z_i \; q(z_i) \,dz_i.\label{eq:zangle}
%& = \sum_{s_i}\int_{x_i} (x_i\cdot s_i) \;q(x_i|s_i)\,q(s_i)\, dx_i,\\
%& = m_{x_i}\!(s_i=1)\,q(s_i=1).
\end{align}
we have
\begin{align}
q(\thev)
%& \propto p(\thev)\;\exp\left(\sum_n\frac{2}{\sigma^2} \Re\left\{y_n\left(\sum_{i=1}^M \langle z_i\rangle^* d_{n i}^*\right) e^{-j\theta_n}\right\}\right),\\
%& \propto p(\thev)\;\prod_n\exp\left(\frac{2 |y_n^*\langle z_n\rangle |}{\sigma^2}\cos(\arg(y_n \sum_{i=1}^M \langle z_i\rangle^* d_{n i}^*)-\theta_n)\right),
& \propto p(\thev)\;\exp\left(\sum_n\frac{2}{\sigma^2} \Re\left\{\eta_n e^{-j\theta_n}\right\}\right),\\
& \propto p(\thev)\;\prod_n\exp\left(\frac{2 |\eta_n|}{\sigma^2}\cos(\arg(\eta_n)-\theta_n)\right).\label{eq:beforapprox}
\end{align}

To go further into the clarification of the procedure, we resort to the following approximation, equalizing a Von Mises \cite{opac-b1101626} and a Gaussian distributions, for small values of $a \in \Rr^+$:
\begin{align}
\frac{1}{2\pi I_0(1/a)} \exp\left(\frac{1}{a}\cos(x-y)\right)\simeq \frac{1}{\sqrt{2\pi a}}\exp\left(-\frac{1}{2a}(x-y)^2\right), \label{eq:approx}
\end{align}
where $I_0(.)$ stands for the modified Bessel of the first kind of order 0.

Under this condition, \eqref{eq:beforapprox} can be rewritten as
\begin{align}
q(\tv)
& \propto p(\thev)\; \prod_n\exp\left(-\frac{ |\eta_n|}{\sigma^2}\left(\theta_n-\arg(\eta_n)\right)^2\right),\\
& \propto p(\thev)\;\exp\left(-\frac{1}{2}(\thev-\arg(\etav))^T\Lambda^{-1}(\thev-\arg(\etav))\right),
\end{align}
with $\etav\triangleq[\eta_1,\ldots,\eta_N]^T$ and $\Lambda^{-1} = \diag\left(\frac{2}{\sigma^2}|\etav|\right)$.

Finally, replacing $p(\tv)$ by its expression \eqref{eq:phaseprior}, we can write the estimate of $q(\tv)$ as a Gaussian distribution as
\begin{align}
%q(\theta_n)
%& \propto \mathcal{N}(m_{\theta_n},\Sigma_{\theta_n}), \label{eq:qtheta}\\
q(\tv) &= \mathcal{N}(\mv_\theta,\Sigma_\theta),\label{eq:qtheta}\\
%\end{align}
%\begin{align}
\text{where }\quad\quad\quad\Sigma_\theta^{-1} &=\Lambda_\theta^{-1} +  \Lambda^{-1},\\
%\begin{pmatrix}
%\frac{\sigma^2(\frac{\sigma_\theta^2}{\sigma_1^2}+a^2)+2\sigma_\theta^2|y_1\langle z_1\rangle^*|}{\sigma^2\sigma_\theta^2} & -\frac{a}{\sigma_\theta^2} & 0 & 0\\
%-\frac{a}{\sigma_\theta^2} &\frac{2\sigma_\theta^2 |y_n^*\langle z_n\rangle |+\sigma^2(1+a^2)}{\sigma^2 \sigma_\theta^2 }&\ddots & 0\\
%0&\ddots & \ddots &-\frac{a}{\sigma_\theta^2} \\
%0 &0 & -\frac{a}{\sigma_\theta^2} & \frac{2\sigma_\theta^2 |y_n^*\langle z_n\rangle |+\sigma^2(1+a^2)}{\sigma^2 \sigma_\theta^2 }\frac{1}{\sigma_\theta^2}
%\end{pmatrix}\\
%\mv_\theta 
%%& = \Sigma_\theta(\Lambda^{-1}\arg(\etav)),\\
%& = \frac{2}{\sigma^2} \Sigma_\theta \etav,
\mv_\theta & = \Sigma_\theta\left( \Lambda^{-1}\arg(\etav)\right),\label{eq:mtheta}
%\Sigma_{\theta_n} \negmedspace&=\negmedspace
%\left\{
%\begin{array}{ll}
%\negmedspace\frac{\sigma^2 \sigma_\theta^2 }{2\sigma_\theta^2 |y_n^*\langle z_n\rangle |+\sigma^2(1+a^2)}&\negmedspace \forall n\in\lbrace 2,\ldots,N\rbrace,\\
%\negmedspace\frac{\sigma^2 \sigma_\theta^2 }{2\sigma_\theta^2 |y_n^*\langle z_n\rangle |+\sigma^2(\frac{\sigma_\theta^2}{\sigma_1^2}+a^2 )}&\negmedspace \text{for } n=1,
%\end{array}
%\right.
\end{align} 
and $\Lambda_\theta^{-1}$ is the precision matrix attached to the prior distribution \eqref{eq:phaseprior} on $\tv$, \ie
\begin{align}
\Lambda_\theta^{-1} &=%\frac{1}{\sigma_\theta^2} 
\begin{pmatrix}
\frac{1}{\sigma_1^2}+\frac{a^2}{\sigma_\theta^2} & -\frac{a}{\sigma_\theta^2} & 0 & 0\\
-\frac{a}{\sigma_\theta^2} & \frac{1+a^2}{\sigma_\theta^2} &\ddots & 0\\
0&\ddots & \ddots &-\frac{a}{\sigma_\theta^2} \\
0 &0 & -\frac{a}{\sigma_\theta^2} & \frac{1}{\sigma_\theta^2}
\end{pmatrix}.\label{eq:lambdatheta}
%m_{\theta_n} = 
%\frac{2\sigma_\theta^2 y_n\langle z_n\rangle^*+ a\sigma^2(m_{\theta_{n+1}}+ m_{\theta_{n-1}})}{4\sigma_\theta^2 |y_n^*\langle z_n\rangle | +2\sigma^2(\frac{\sigma_\theta^2}{\sigma_1^2}+a^2)},\label{eq:mtheta}%{\sigma^2\sigma_\theta^2},
\end{align}
%Vector $\etav$ is defined as $\etav\triangleq[\eta_1,\ldots,\eta_N]^T$, with
%\begin{align}
%%\eta_n = y_n\sum_i q(\!s_i=1\!)\; m_{x_i}\!(s_i=1)^*d_{ni}^*,
%\eta_n = y_n\sum_{i=1}^M \langle z_i\rangle^*d_{ni}^*,
%\end{align}
%where $d_{ni}^*$ is the conjugate of the $n$th element of $\dv_i$,  and $\langle z_i\rangle$ is defined using the current estimate of $q(z_i)$ as
%% \suppCH{the current estimates $q(s_i)$ and} $m_{x_i}\!(s_i) = \int_{x_i} x_i\, q(x_i|s_i)$. \remCH{$q(s_i)$ et $q(x_i|s_i)$ non definis a ce stade: a deplacer plus haut?}
%\begin{align}
% \langle z_i\rangle
%&\triangleq \int_{z_i} z_i \; q(z_i) \,dz_i.\label{eq:zangle}
%%& = \sum_{s_i}\int_{x_i} (x_i\cdot s_i) \;q(x_i|s_i)\,q(s_i)\, dx_i,\\
%%& = m_{x_i}\!(s_i=1)\,q(s_i=1).
%\end{align}

Note that the distribution $q(\tv)$ being Gaussian, the marginals $q(\theta_n)$ come straightforwardly as
\begin{align}
q(\theta_n) = \mathcal{N}(m_{\theta_n}, \Sigma_{\theta_n}), 
\end{align}
where $m_{\theta_n}$ (resp. $\Sigma_{\theta_n})$) is the $n$th element in $\mv_\theta$ (resp. in the diagonal of $\Sigma_\theta$). In practice, estimating 
%the parameters $m_{\theta_n}$ and $\Sigma_{\theta_n}$ of the marginal distributions $q(\theta_n)$ 
these parameters can be efficiently implemented through a Kalman smoother \cite{Mendel_book} because of the particular structure of the precision matrix \eqref{eq:lambdatheta}.

\subsection{Update of $q(z_i)$} %$q(\xv,\sv)$}
\label{subsec:algo}
%The iterates of the
%%$q(x_i,s_i)$ 
%$q(z_i)$'s 
%take the form of a mixture of two Gaussian distributions and only depend on the marginal distributions $q(\theta_n)$ of $q(\tv)$. %More precisly, recalling that $$q(z_i) = q(x_i,s_i), $$ we obtain%= q(x_i|s_i)\,q(s_i),$$ we obtain
%From an implementation point of view, the parameters $\mv_{\theta}$ and $\Sigma_{\theta}$ are sufficient to entirely characterize the Gaussian distribution \eqref{eq:qtheta}. In the update of the posterior distribution They are then used in the update of the posterior distribution $q(x_i,s_i)$ as follows \remCH{mentionner que $q(x_i,s_i)=q(x_i|s_i)q(s_i)$}
The iterates of the $q(z_i)$'s are written as
\begin{align}
%&q(x_i,s_i)\nonumber\\
q(z_i) %\nonumber\\
&\propto \exp\left(\int_{\thev} q(\thev)\int_{\zv_{\neq i}}\prod_{k\neq i} q(z_k) \log p(\yv,\zv,\thev)\;d\thev\; d\zv\right),\\
&\propto p(x_i)\;p(s_i)\;\exp\left(\int_{\thev} q(\thev)\int_{\zv_{\neq i}} \prod_{k\neq i}q(z_k) \log p(\yv|\zv,\thev)\;d\thev\; d\zv\right),\\
&\propto p(x_i)\;p(s_i)\\
%&\exp\left(-\frac{1}{\sigma^2} \left( 
%s_i x_i \sum_{k\neq i} \left< s_k x_k^*\right> \dv_{k}^H\dv_{i}
%+s_i x_i^* \sum_{k\neq i} \langle s_k x_k\rangle \dv_{i}^H\dv_{k}
%+ s_i x_i^* x_i \;\dv_i^H\dv_i-2\; \sum_n \langle \Re(y_n^* s_i x_i d_{n i}e^{j\theta_n}) \rangle \right)\right), \label{eq:q(xs)}
&\exp\left(-\frac{1}{\sigma^2} \left( 
z_i \sum_{k\neq i} \left< z_k\right>^* \dv_{k}^H\dv_{i}
+z_i^* \sum_{k\neq i} \langle z_k\rangle \dv_{i}^H\dv_{k}
+ z_i^*z_i \;\dv_i^H\dv_i-2\; \sum_n \Re\bigg\{y_n^* z_i d_{n i}\langle e^{j\theta_n} \rangle\bigg\} \right)\right),\nonumber\\
&\propto \exp\left(-\frac{1}{\sigma^2}\left(2\;\Re\bigg\{z_i\sum_{k\neq i} \langle z_k\rangle^* \dv_{k}^H\dv_{i}\bigg\}+ z_i^*z_i \;\dv_i^H\dv_i-2\; \sum_n  \Re\bigg\{y_n^* z_i d_{n i}\langle e^{j\theta_n}\rangle\bigg\} \right)\right), \label{eq:q(xs)}
\end{align}
with $\langle z_k\rangle$ defined in \eqref{eq:zangle}
and
\begin{align}
\langle e^{j\theta_n} \rangle
&\triangleq \int_{\theta_n} e^{j\theta_n}q(\theta_n)\;d\theta_n,\\
&=\int_{\theta_n} e^{j\theta_n}\mathcal{N}(m_{\theta_n},\Sigma_{\theta_n})\;d\theta_n,\\
&=\frac{1}{\sqrt{2\pi\Sigma_{\theta_n}}} \int_{\theta_n} e^{j\theta_n}\exp(-\frac{1}{2\Sigma_{\theta_n}}(\theta_n-m_{\theta_n})^2)\;d\theta_n,\\
&\simeq \frac{1}{2\pi I_0(1/\Sigma_{\theta_n})}  \int_{\theta_n} e^{j\theta_n}\exp\left(\frac{1}{\Sigma_{\theta_n}}\cos(\theta_n-m_{\theta_n})\right)\;d\theta_n,\\
&\simeq \frac{1}{2\pi I_0(1/\Sigma_{\theta_n})}  \int_{\theta_n} \cos(\theta_n)\exp\left(\frac{1}{\Sigma_{\theta_n}}\cos(\theta_n-m_{\theta_n})\right)\;d\theta_n \nonumber\\
&\quad\quad+ j \frac{1}{2\pi I_0(1/\Sigma_{\theta_n})}  \int_{\theta_n} \sin(\theta_n)\exp\left(\frac{1}{\Sigma_{\theta_n}}\cos(\theta_n-m_{\theta_n})\right)\;d\theta_n,\\
&\simeq \frac{1}{2\pi I_0(1/\Sigma_{\theta_n})}  \int_{\theta_n} \cos(\theta_n+m_{\theta_n})\exp\left(\frac{1}{\Sigma_{\theta_n}}\cos(\theta_n)\right)\;d\theta_n \nonumber\\
&\quad\quad+ j \frac{1}{2\pi I_0(1/\Sigma_{\theta_n})}  \int_{\theta_n} \sin(\theta_n+m_{\theta_n})\exp\left(\frac{1}{\Sigma_{\theta_n}}\cos(\theta_n)\right)\;d\theta_n,\\
&\simeq \frac{1}{2\pi I_0(1/\Sigma_{\theta_n})} \cos(m_{\theta_n}) \int_{\theta_n} \cos(\theta_n)\exp\left(\frac{1}{\Sigma_{\theta_n}}\cos(\theta_n)\right)\;d\theta_n \nonumber\\
&\quad\quad-\frac{1}{2\pi I_0(1/\Sigma_{\theta_n})} \sin(m_{\theta_n}) \int_{\theta_n} \sin(\theta_n)\exp\left(\frac{1}{\Sigma_{\theta_n}}\cos(\theta_n)\right)\;d\theta_n \nonumber\\
&\quad\quad+ j \frac{1}{2\pi I_0(1/\Sigma_{\theta_n})} \sin(m_{\theta_n}) \int_{\theta_n} \cos(\theta_n)\exp\left(\frac{1}{\Sigma_{\theta_n}}\cos(\theta_n)\right)\;d\theta_n\\
&\quad\quad+ j \frac{1}{2\pi I_0(1/\Sigma_{\theta_n})}  \cos(m_{\theta_n}) \int_{\theta_n} \sin(\theta_n)\exp\left(\frac{1}{\Sigma_{\theta_n}}\cos(\theta_n)\right)\;d\theta_n,\\
&\simeq \frac{1}{2\pi I_0(1/\Sigma_{\theta_n})} \cos(m_{\theta_n}) \int_{\theta_n} \cos(\theta_n)\exp\left(\frac{1}{\Sigma_{\theta_n}}\cos(\theta_n)\right)\;d\theta_n \nonumber\\
&\quad\quad+ j \frac{1}{2\pi I_0(1/\Sigma_{\theta_n})} \sin(m_{\theta_n}) \int_{\theta_n} \cos(\theta_n)\exp\left(\frac{1}{\Sigma_{\theta_n}}\cos(\theta_n)\right)\;d\theta_n,\\
&\simeq \frac{I_1(1/\Sigma_{\theta_n})}{I_0(1/\Sigma_{\theta_n})} e^{jm_{\theta_n}}, \label{eq:qtheta}
\end{align}
where we have used the approximation \eqref{eq:approx} and $I_1(.)$ stands for the modified Bessel of the first kind of order 1.

Coming back to \eqref{eq:q(xs)} and recalling that $$q(z_i) = q(x_i,s_i), $$ we finally obtain
\begin{align}
q(x_i|s_i)&=\mathcal{C}\Ns(m_{x_i}\!(s_i),\Sigma_{x_i}\!(s_i)),\label{eq:qxisi2}\\
q(s_i) & \propto \sqrt{\Sigma_{x_i}\!(s_i)} \exp \left (\frac{m_{x_i}\!(s_i)^*m_{x_i}\!(s_i)}{\Sigma_{x_i}\!(s_i)} \right )p(s_i),\label{eq:qsi2}
\end{align}
where $\propto$ means proportionality, 
\begin{align}
	&\Sigma_{x_i}\!(s_i)=\frac{\sigma^2\sigma^2_x}{\sigma^2+ s_i \sigma_x^2\dv_i^H\dv_i}, \\
	&m_{x_i}\!(s_i) = s_i\frac{\sigma_x^2}{\sigma^2+ s_i \sigma_x^2\dv_i^H\dv_i} \dv_i^H\langle\rv_i\rangle, \label{eq:moy2}\\
	&\langle \rv_i \rangle=\bar{\yv}-\sum_{k\neq i}q(s_k=1)\, m_{x_k}\!(s_k=1)\,\dv_k,\label{eq:resi2}\\
	&\bar{\yv}=\left [y_n e^{-j m_{\theta_n}}\;\frac{ I_1(1/\Sigma_{\theta_n})}{I_0(1/\Sigma_{\theta_n})}\right]_{n=\lbrace1\ldots M\rbrace}.\label{eq:endproc}
\end{align}
% and $I_0$ (resp. $I_1$) stands for the modified Bessel of the first kind of order 0 (resp. 1).
%Note that, in practice, estimating the parameters $m_{\theta_n}$ and $\Sigma_{\theta_n}$ of the marginal distributions $q(\theta_n)$ can be efficiently implemented through a Kalman smoother \cite{Mendel_book}.
Within the above notations, expression \eqref{eq:zangle} simply writes as $\langle z_i \rangle= q(s_i=1)\; m_{x_i}\!(s_i=1)$, as well as the estimates $\hat{z_i}$ in \eqref{eq:estiz}, using the final estimates of $q(s_i)$ and $m_{x_i}\!(s_i)$, after convergence of the algorithm.

We note that the update equation \eqref{eq:qxisi2}-\eqref{eq:endproc} share some connections with the phase retrieval algorithm presented in \cite{Dremeau2015}.
The latter, relying also on a VBEM algorithm, differs from the proposed procedure in the definition of the prior distributions \eqref{eq:phaseprior} and \eqref{eq:priors}, respectively replaced by a uniform and a Gaussian distributions. 
%The latter relies also on a VBEM algorithm, and shares a similar structure as the algorithm proposed here. 
In practice, both procedures share a similar structure. 
Leaving out 
%the difference attached to 
the choice made here of a sparse-enforcing prior on $\zv$, the main difference lies in the ``reconstructed'' phases $m_{\theta_n}$ in \eqref{eq:endproc}: while their definition relies here on the parameters of the Markov chain through the precision matrix \eqref{eq:lambdatheta}, they only depend on the observations in \cite{Dremeau2015} where a non-informative prior is considered.
%while their definitions are linked to one another here, they only consider the first term attached to the observations in \cite{Dremeau2015}.
%linked to one another by \eqref{eq:mtheta} while they only consider the first term attached to the observation in \cite{}.

%Coming back to problem \eqref{eq:pb}, $p(\zv|\yv)$ is simplified as $p(\zv|\yv)\simeq\int_\tv q(\tv)\prod_i q(z_i) = \prod_i q(z_i)$, after convergence of the procedure. Then the MMSE estimate of $\hat{\zv}$ comes from the element-wise operation: $\forall i\in \lbrace 1,\ldots,M\rbrace$, $\hat{z}_i = \langle z_i\rangle$ defined in \eqref{eq:zangle} using the final estimate of $q(z_i)$ as
%\begin{align}
%\hat{z}_i
%%&=\int_{z_i} z_i \; q(z_i) \,dz_i,\\
%& = \sum_{s_i}\int_{x_i} (x_i\cdot s_i) \;q(x_i|s_i)\,q(s_i)\, dx_i,\\
%& = m_{x_i}\!(s_i=1)\,q(s_i=1).
%\end{align}

\subsection{Noise estimation}
\label{subsec:noiseest}
As emphasized in \cite{Dremeau2012}, the estimation of model parameters can easily be embedded within the VBEM procedure. Among them, the noise variance is of particular interest. Measure of the (mean) discrepancies between the observations and the assumed model, its iterative estimation usually helps the convergence of the algorithm to a proper local minimum, as observed in \cite{Dremeau2012}. Particularized to model \eqref{eq:obs}-\eqref{eq:priors}, this leads to
\begin{align}
\hat{\sigma}^2\negmedspace
&= \argmax_{\sigma^2}\negmedspace \int_{\zv}q(\zv,\tv)\log p(\yv,\zv, \tv ; \sigma^2)\,  d\zv\, d\tv,\nonumber\\
&= \argmax_{\sigma^2}\nonumber\\
		&\quad\negmedspace \int_{\zv}q(\zv,\tv)\bigg[-N\log (\sigma^2)-\frac{1}{\sigma^2}\bigg(\yv^H\yv-2\sum_i \sum_n \Re\bigg\{y_n^* z_i d_{n i} e^{j\theta_n}\bigg\}+\sum_i\sum_{k} s_i\, s_k\,x_i^*\,x_k\,\dv_i^H\dv_k \bigg)\bigg]\,  d\zv\, d\tv,\nonumber\\
%&= \argmax_{\sigma^2}\nonumber\\
%&\quad\negmedspace\bigg[-N\log (\sigma^2)-\frac{1}{\sigma^2}\bigg(\yv^H\yv-2\sum_i \sum_n \Re\bigg\{y_n^* \langle s_i\,x_i\rangle_{q(x_i,s_i)} d_{n i} \langle e^{j\theta_n}\rangle\bigg\}-2\sum_i\Re\bigg\{\langle s_i\,x_i\rangle_{q(x_i,s_i)}\,\bar{\yv}^H\dv_i\bigg\}+\sum_i\sum_{k} \langle s_i\, s_k\,x_i^*\,x_k\rangle_{q(x_i,s_i)\,q(x_k,s_k)}\,\dv_i^H\dv_k \bigg)\bigg],\nonumber
&= \argmax_{\sigma^2}\nonumber\\
&\quad\negmedspace\bigg[-N\log (\sigma^2)-\frac{1}{\sigma^2}\bigg(\yv^H\yv-2\sum_i \sum_n \Re\bigg\{y_n^* \langle s_i\,x_i\rangle d_{n i} \langle e^{j\theta_n}\rangle\bigg\}+\sum_i\sum_{k} \langle s_i\, s_k\,x_i^*\,x_k\rangle\,\dv_i^H\dv_k \bigg)\bigg],\nonumber\\ %-2\sum_i\Re\bigg\{\langle s_i\,x_i\rangle\,\bar{\yv}^H\dv_i\bigg\}
&= \argmax_{\sigma^2}\nonumber\\
&\quad\negmedspace\bigg[-N\log (\sigma^2)-\frac{1}{\sigma^2}\bigg(\yv^H\yv-2\sum_i\Re\bigg\{\langle s_i\,x_i\rangle\,\bar{\yv}^H\dv_i\bigg\}+\sum_i\sum_{k} \langle s_i\, s_k\,x_i^*\,x_k\rangle\,\dv_i^H\dv_k \bigg)\bigg],\nonumber 
\end{align}
where $\langle e^{j\theta_n}\rangle$ is defined as in \eqref{eq:qtheta}, $\bar{\yv}$ as in \eqref{eq:endproc} and
\begin{align}
\langle s_i\,x_i\rangle%_{q(x_i,s_i)} 
&=\sum_{s_i\in\lbrace 0,1\rbrace}\int_{x_i} s_i\, x_i\, q(x_i,s_i)\,dx_i\nonumber\\
&=\sum_{s_i\in\lbrace 0,1\rbrace}s_i\, q(s_i)\int_{x_i}  x_i\, q(x_i|s_i)\,dx_i\nonumber\\
&=q(s_i=1)\,\int_{x_i} x_i\, q(x_i|s_i=1)\,dx_i\nonumber\\
&=q(s_i=1)\,m_{x_i}(s_i=1)\nonumber\\
 \langle s_i\, s_k\,x_i^*\,x_k\rangle%_{q(x_i,s_i)\,q(x_k,s_k)}
 &=\sum_{s_i\in\lbrace 0,1\rbrace}\sum_{s_k\in\lbrace 0,1\rbrace}\int_{x_i}\int_{x_k} s_i\,s_k\, x_i^*x_k\, q(x_i,s_i)\, q(x_k,s_k)\,dx_i\,dx_k\nonumber\\
&=\sum_{s_i\in\lbrace 0,1\rbrace}\negmedspace s_i\, q(s_i)\sum_{s_k\in\lbrace 0,1\rbrace}\negmedspace s_k\,q(s_k)\int_{x_i} \negmedspace x_i^*q(x_i|s_i)\, dx_i\,\int_{x_k}\negmedspace x_k\,q(x_k|s_k)\,dx_k\quad \text{if } k\neq i,\nonumber\\
&=q(s_i=1)\,q(s_k=1)\,m_{x_i}(s_i=1)^*m_{x_k}(s_k=1)\quad \quad\quad\quad\quad\quad\quad\quad\quad\quad\text{if } k\neq i,\nonumber\\
&=\sum_{s_i\in\lbrace 0,1\rbrace}\negmedspace s_i^2\, q(s_i)\int_{x_i} \negmedspace x_i^*x_i\,q(x_i|s_i)\, dx_i\;\;\;\quad\quad\quad\quad\quad\quad\quad \quad\quad\quad\quad\quad\quad\quad\text{if } k= i,\nonumber\\
&=q(s_i=1)\bigg(\Sigma_{x_i}(s_i=1)+ m_{x_i}(s_i=1)^*\;m_{x_i}(s_i=1)\bigg)\quad\quad\quad\quad \quad\quad\quad\text{if } k= i.\nonumber
\end{align}
%	\hat{\sigma}_n^2&=\argmin_{\var_n} \mathrm{KL}(q(\xv,\sv);p(\xv,\sv,\yv)),\label{eq:estimvar}\\
%	         &=\frac{1}{M} \left \langle  \| \yv -  \exp^{j\theta}\sum_i  x_i \dv_i \|^2 \right  \rangle_{\prod_i q(x_i,s_i)}\\
Then, derivating and setting the resulting expression to zero, we get:
\begin{align}
\frac{N}{\hat{\sigma}^2}
&=\frac{1}{\hat{\sigma}^4}\bigg(\yv^H\yv-2\sum_i\Re\bigg\{\langle s_i\,x_i\rangle\,\bar{\yv}^H\dv_i\bigg\}+\sum_i\sum_{k} \langle s_i\, s_k\,x_i^*\,x_k\rangle\,\dv_i^H\dv_k \bigg),\nonumber\\
\hat{\sigma}^2\negmedspace
	         &=\frac{1}{N} \left(\yv^H\yv-2\sum_i\Re\bigg\{q(s_i=1)\,m_{x_i}(s_i=1)\,\bar{\yv}^H\dv_i\bigg\} \right. \nonumber\\
	         &\quad+\sum_i\sum_{k\neq i} q(s_i=1)\; q(s_k=1)\; m_{x_i}(s_i=1)^*\,m_{x_k}(s_k=1)\;\dv_i^H\dv_k \nonumber\\
	         &\quad\left.+\sum_i q(s_i=1)\,\bigg(\Sigma_{x_i}(s_i=1)+ m_{x_i}(s_i=1)^*\,m_{x_i}(s_i=1)\bigg)\;\dv_i^H\dv_i\right).
\end{align}
In the following, we will refer to the proposed procedure as ``paVBEM'' for ``phase-aware VBEM algorithm".

\section{Experiments}
\label{sec:expe}

In this section, we assess numerically the effectiveness of the proposed approach. We consider the problem of identifying the directions of arrival of $K$ plane waves from $N=256$ observations ($K$ will be specified later on). We assume that the angles of the $K$ incident waves can be written as $\phi_k =- \frac{\pi}{2} +  i_k  \frac{\pi}{50}$ with $i_k\in[1, 50]$, $\forall k\in\lbrace1,\ldots,K\rbrace$. The set of angles $\{\phi_i = - \pi +  i  \frac{\pi}{50}\}_{i\in\lbrace1,\ldots,50\rbrace}$ together with the choice of the parameter $\Delta/\lambda=4$ define the columns of the dictionary $\D$ (see section \ref{sec:bayes}). We set the following parameters for the phase Markov model \eqref{eq:phaseprior}: $\sigma_1^2=10^6$, $\sigma_\theta^2=1$ and $a=0.8$. This corresponds to the situation where one has a large uncertainty on the initial value of the phase noise but connections exist between the phase noise on adjacent sensors. 

%We assess the performance of the DOA procedures by considering the normalized correlation between the ground truth $\sv$ and its reconstruction $\hat{\sv}$, \ie $\frac{\vert \sv^H\hat{\sv}\vert}{\|\sv\|\|\hat{\sv}\|}$ \remCH{Il faut sans doute dire quelque part comment on calcule notre $\hat{\sv}$?}. 
We consider the normalized correlation between the ground truth $\zv$ and its reconstruction $\hat{\zv}$, \ie $\frac{\vert \zv^H\hat{\zv}\vert}{\|\zv\|_2\|\hat{\zv}\|_2}$, as a figure of merit. 
This quantity is averaged over $50$ realizations for each point of simulation.  We compare the performance of the following algorithms: \textit{i)} the standard beamforming introduced in \cite{Johnson1993} (blue curve, circle mark); \textit{ii)} the so-called \textit{prVBEM} algorithm proposed in \cite{Dremeau2015} as a solution to the phase retrieval problem (red curve, triangle mark); \textit{iii)} the \textit{paVBEM} procedure described in section \ref{sec:procedure} (magenta curve, diamond mark); \textit{iv)} a relaxed version of \textit{paVBEM} in which the sparsity of $\zv$ is not exploited but replaced by a Gaussian prior (cyan curve, square mark). 

%\remCH{The proposed procedure as well as the methodologies exposed in \cite{Dremeau2015} and \addCH{\cite{aaaa}} are until convergence. }\remCH{Mentionne-t-on leur initialisation?}
%\begin{itemize}
%\item[\textbullet] the standard beamforming introduced in \cite{Johnson1993} (blue curve).
%\item[\textbullet] the ``phase retrieval'' algorithm proposed in \cite{Dremeau2015} (red curve). 
%\item the proposed procedure (magenta curve).
%\end{itemize}

\pgfplotsset{
ymin=0,
ymax=1,
xlabel={$\sigma^2$},
ylabel={$\frac{\vert \hat{\mathbf{z}}^H \mathbf{z} \vert}{\|\mathbf{z}\|_2 \|\mathbf{z}\|_2}$},
grid=major,
legend columns=1,
legend style={font=\tiny}
}

\begin{figure}[t!]
\begin{center}
%\begin{kfig}{2}

\begin{tikzpicture}[font=\large]%[trim axis left, trim axis right]
\begin{semilogxaxis}[
mark size = 1.5,
%title   ={$k=2$},
legend style={at={(0.02,0.15)},anchor=west},
line width = 0.6,
scale = 0.9
]
\addplot[color=blue,mark=o,solid] table[x index=0, y index=1] {./Corr_PR_noise_variance_dimy256_dims50_k2_ntrial50_varx1_varn0.01_varp11000000_varpt1_DoA_MarkovRand.dat};
%%\addlegendentry{Pseudo inverse}
\addplot[color=red,mark size = 2, mark=triangle,solid] table[x index=0, y index=2] {./Corr_PR_noise_variance_dimy256_dims50_k2_ntrial50_varx1_varn0.01_varp11000000_varpt1_DoA_MarkovRand.dat};
%%\addlegendentry{VBEM uniform}
\addplot[color=cyan,mark=square,solid] table[x index=0, y index=3] {./Corr_PR_noise_variance_dimy256_dims50_k2_ntrial50_varx1_varn0.01_varp11000000_varpt1_DoA_MarkovRand.dat};
%%\addlegendentry{VBEM non-uniform}
%\addplot[color=red,mark=o,dashed] table[x index=0, y index=4] {../Results/Corr_PR_noise_variance_dimy256_dims50_k2_ntrial50_varx1_varn0.01_varp11000000_varpt1_DoA_MarkovRand.dat};
%%%\addlegendentry{VBEM uniform BG}
\addplot[color=magenta,mark size = 2, mark=diamond,solid] table[x index=0, y index=7] {./Corr_PR_noise_variance_dimy256_dims50_k2_ntrial50_varx1_varn0.01_varp11000000_varpt1_DoA_MarkovRand.dat};
%%%\addlegendentry{VBEM non-uniform BG}
\end{semilogxaxis}
\end{tikzpicture}
\end{center}\vspace{-0.6cm}
\caption{Evolution of the (averaged) normalized correlation as a function of the variance $\sigma^2$  when $K=2$.\label{fig:k=2}}
\end{figure}
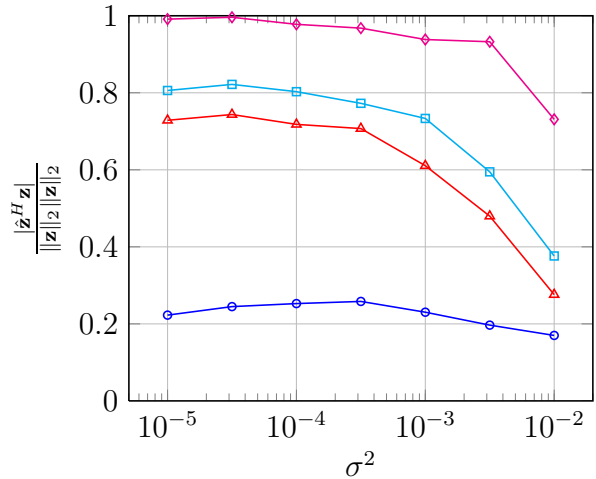

\begin{figure}[t!]
\begin{center}
%\begin{kfig}{2}

\begin{tikzpicture}[font=\large]%[trim axis left, trim axis right]
\begin{semilogxaxis}[
mark size = 1.3,
%title   ={$k=5$},
legend style={at={(0.02,0.15)},anchor=west},
line width = 0.6,
scale = 0.9
]
\addplot[color=blue,mark=o,solid] table[x index=0, y index=1] {./Corr_PR_noise_variance_dimy256_dims50_k5_ntrial50_varx1_varn0.01_varp11000000_varpt1_DoA_MarkovRand.dat};
%%\addlegendentry{Pseudo inverse}
\addplot[color=red,mark size = 2,mark=triangle,solid] table[x index=0, y index=2] {./Corr_PR_noise_variance_dimy256_dims50_k5_ntrial50_varx1_varn0.01_varp11000000_varpt1_DoA_MarkovRand.dat};
%%\addlegendentry{VBEM uniform}
\addplot[color=cyan,mark=square] table[x index=0, y index=3] {./Corr_PR_noise_variance_dimy256_dims50_k5_ntrial50_varx1_varn0.01_varp11000000_varpt1_DoA_MarkovRand.dat};
%%\addlegendentry{VBEM non-uniform}
%\addplot[color=red,mark=o,dashed] table[x index=0, y index=4] {../Results/Corr_PR_noise_variance_dimy256_dims50_k5_ntrial50_varx1_varn0.01_varp11000000_varpt1_DoA_MarkovRand.dat};
%%%\addlegendentry{VBEM uniform BG}
\addplot[color=magenta,mark size = 2,mark=diamond] table[x index=0, y index=7] {./Corr_PR_noise_variance_dimy256_dims50_k5_ntrial50_varx1_varn0.01_varp11000000_varpt1_DoA_MarkovRand.dat};
%%%\addlegendentry{VBEM non-uniform BG}
\end{semilogxaxis}
\end{tikzpicture}

\end{center}\vspace{-0.6cm}
\caption{Evolution of the (averaged) normalized correlation as a function of the variance $\sigma^2$  when $K=5$.\label{fig:k=5}}
\end{figure}
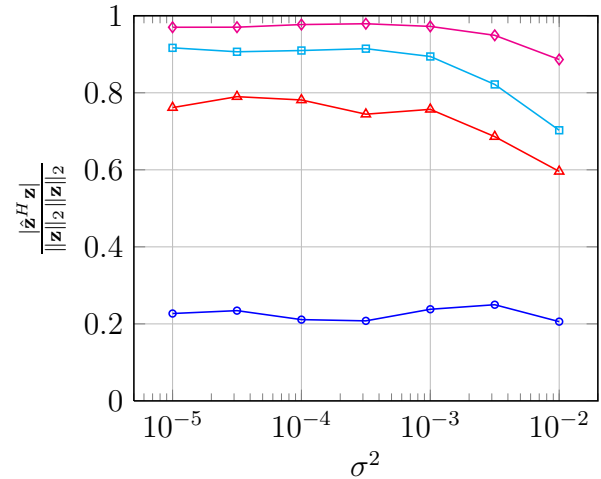

 The performance of these procedures are illustrated in Fig.~\ref{fig:k=2}  and~\ref{fig:k=5} as a function of the noise variance $\sigma^2$ for $K=2$ and $K=5$, respectively. We see that the beamforming algorithm, which was originally  proposed to solve the DOA estimation problem in the standard linear setup ($\Pn=\I_N$),  fails to cope with the presence of fluctuations in the phase $\boldsymbol{\theta}$. The three other algorithms achieve different levels of performance, depending on the power of the additive noise and the number of incident waves. We note that all these procedures are derived from a similar optimization procedure but consider different degrees of knowledge on $\boldsymbol{\theta}$ and $\zv$. In \cite{Dremeau2015}, the authors assume that the phase is uniformly distributed and the sparse nature of $\zv$ is ignored; in the relaxed version of the proposed procedure the phase model \eqref{eq:phaseprior} is exploited but the sparsity of $\zv$ is not taken into account; finally, as explained previously, the methodology presented in this paper integrates both the phase model \eqref{eq:phaseprior} and the sparsity of $\zv$ in the estimation process. We see from Fig.~\ref{fig:k=2} and~\ref{fig:k=5}  that the performance of these algorithms directly relates to the level of information they exploit: the proposed methodology outperforms its relaxed counterpart which, in turn, leads to better performance than the procedure proposed in \cite{Dremeau2015}. 

We also notice that the procedures achieve better performance when $K=5$ than $K=2$. This counter-intuitive behavior is typical for phase-retrieval problems. In fact, it is easy to see that, when no phase information is available (\ie $\theta_n$ is uniform on $[0,2\pi]$), only the modulus of $y_n$ provides some information on $\zv$. In such a case, the worst situation occurs when $K=1$ since all the elements of $\D$ has the same modulus (equal to 1) and the observations thus provide no information on $\zv$. The setup $K=2$ is close to this worst case, hence explaining the observed behavior. 

%In fact, the worst case occurs for $K=1$ when the algorithms have no information about the phase (as \eg in \cite{Dremeau2015}) because all the elements of $\D$ has the same modulus but different phase. %, the worst case 
%Similarly to the proposed procedure, this algorithm relies on a mean-field approximation of a maximum a posteriori problem but, unlike in this paper, the sparsity of $\sv$ and the dependence between the elements of $\boldsymbol{\theta}$ are taken into account. 

\section{Conclusion}
We have presented a novel algorithm able to estimate the direction of arrivals of plane waves in environments corrupted by phase noise. Inspired by recent works in phase retrieval \cite{Schniter2012, Dremeau2015}, our approach relies on a mean-field approximation and exploits two types of priors: on the DOA through a sparse-enforcing distribution as in \cite{Xenaki2014}, and on the phase noise through a Markov model.  Our experiments have confronted the proposed approach %- named \textit{paVBEM} - 
to conventional beamforming and similar variational approaches handicapped by non-informative priors. In this regard,
%\textit{paVBEM} 
its good performance tends to prove a successful inclusion of the priors. 
Future work will include further assessment in underwater acoustics.
%of \textit{paVBEM}, in particular 
%
%In particular, we intend to test its performance on DOA estimation of underwater sources.

%In the experimental phase of our contribution, we To assess the performance of the algorithm, we have proposed experiments the conventional beamforming and similar variational approaches handicapped by non-informative priors. our procedure presents a good performance, proving the 

%\remCH{La police n'est pas la meme pour la section "Acknowledgment". Verifier dans les guidelines comment faire les acknowledgements ? (Enlever la section et mettre le remerciement en footnote)}
%\subsubsection*{Acknowledgment}
%
%This work has been supported by the DGA/MRIS.

\newpage

\bibliographystyle{IEEEbib}
\small

\bibliography{Biblio}

\end{document}